\documentstyle[mprocl]{article}

\def\beq{\begin{equation}} \def\eeq{\end{equation}}  

\def\rightcontract{\mathop{\hbox{\vrule width0.5pt height6pt%
  \vrule height0.5pt width6pt}}}
\def\Lie{\hbox{\it\char'44}}

\def\fl{}

\def\journalfont{\it}         
\def\jou#1{{\journalfont #1\ }}
\def\joudef#1#2{\def #1{\jou{\ignorespaces #2}}}

\joudef{\AAA}{  Astron.\ Astrophys.}
\joudef{\AJP}{  Am.\ J.\ Phys.}
\joudef{\AIP}{  Adv.\ Phys.}
\joudef{\AM}{   Ann.\ Math.}
\joudef{\AP}{   Ann.\ Phys.\ (N.Y.)}
\joudef{\AOP}{   Ann.\ Phys.\ (N.Y.)}
\joudef{\APJ}{  Astrophys.\ J.}
\joudef{\CJP}{  Can.\ J.\ Phys.}
\joudef{\CMP}{  Commun.\ Math.\ Phys.}
\joudef{\CQG}{  Class.\ Quantum Grav.}
\joudef{\GRG}{  Gen.\ Relativ.\ Grav.}
\joudef{\IJMP}{ Int.\ J.\ Mod.\ Phys.}
\joudef{\IJTP}{ Int.\  J.\ Theor.\ Phys.}
\joudef{\JKPS}{ J.\ Korean\ Phys.\ Soc.}
\joudef{\JMP}{  J.\ Math.\ Phys.}
\joudef{\JPAMG}{ J.\ Phys.\ A: Math.\ Gen.}
\joudef{\MNRAS}{ Mon.\ Not.\ R.\ Ast.\ Soc.}
\joudef{\NAT}{  Nature}
\joudef{\NCIM}{ Nuovo Cim.}
\joudef{\NUCP}{ Nuc.\ Phys.}
\joudef{\NCB}{  Il Nuovo Cimento ``B}
\joudef{\PL}{   Phys.\ Lett.}
\joudef{\PR}{   Phys.\ Rev.}
\joudef{\PREP}{ Phys.\ Rep.}
\joudef{\PRL}{  Phys.\ Rev.\ Lett.}
\joudef{\PTP}{  Prog.\ Theor.\ Phys.}
\joudef{\RMP}{  Rev.\ Mod.\ Phys.}
\joudef{\SPJ}{  Sov.\ Phys.\ JETP}

\def\vol#1{{\bf #1}}  \def\book#1{{\it #1\/}}  \def\art#1{{#1}}
\def\journal#1{{\it #1\/}}  \def\pub(#1){(#1)}  \def\ed#1{ed #1}
\begin{document}

\title{The Inertial Forces / Test Particle Motion Game}

\author{Donato Bini}
\address{
  Istituto per Applicazioni della Matematica C.N.R.,\\ 
  I--80131 Napoli, Italy and\\
International Center for Relativistic Astrophysics, University of Rome,\\
 I--00185 Roma, Italy }  

\author{Paolo Carini}
\address{
  International Center for Relativistic Astrophysics,
  University of Rome,\\
 I--00185 Roma, Italy}

\author{Robert T. Jantzen}
\address{
  International Center for Relativistic Astrophysics,
  University of Rome, \\
I--00185 Roma, Italy and  \\
Department of Mathematical Sciences, Villanova University,
  Villanova, \\
PA 19085, USA}

\maketitle\abstracts{
The somewhat fragmented body of current literature analyzing the
properties of test particle motion in static and stationary spacetimes 
and in general spacetimes is pulled together and clarified using the
framework of gravitoelectromagnetism.
}


\section{Introduction}

During the past decade a long list of authors have studied various 
properties of test particle motion in static or stationary 
axisymmetric spacetimes, usually in ``uniform circular motion" along an 
integral curve of a timelike Killing vector field associated with this 
symmetry. Many of these authors have decomposed the acceleration 
vector field along the test world line with respect to a family 
of observers defined on most of the spacetime which are also in 
uniform circular motion. This acceleration vector is orthogonal 
to the cylindrical world sheet symmetry group orbits containing 
the circular world lines, the case of purely transverse relative
acceleration with respect to any of these observer families, and 
so the observer spatial projection acts as the identity on the  
acceleration vector.
Thus in this special case, the relative observer / relative motion
orthogonal splitting $4 \to 3+1 \to (2+1)+1$ of the full tangent 
space first into the local rest space and local time direction of 
the observer and then of the local rest space into its transverse 
and longitudinal subspaces relative to the direction of relative 
motion of the test particle provides a representation of the full 
acceleration vector itself rather than just of its observer  
spatial projection.
However, the relative observer approach has the advantage over 
the direct spacetime approach of 
offering a directly interpretable physical relative velocity 
variable to parametrize the circular motion, whereas the 
spacetime approach is limited to the use of the coordinate
angular velocity variable whose interpretation (as a speed 
to be compared to the speed of light) depends on further 
quantities.

Some authors have attempted to generalize a conformal variation 
of this relative observer / relative motion splitting 
of the acceleration of the acceleration vector of a test particle
from these special symmetry conditions to general motion in general 
spacetimes. However, the general 
language of gravitoelectromagnetism, which provides a systematic 
approach to work much of which is many decades old, already 
provides unambiguously all possible such acceleration decompositions 
in a general spacetime. For a rotating black hole spacetime where
there is a natural slicing by the integral hypersurfaces of the 
local rest spaces of the locally nonrotating (zero angular 
momentum) observers and a natural threading by the world lines of 
the static Killing observers which are at rest with respect to 
spatial infinity, there are three different descriptions 
(hypersurface, threading, and slicing) and 3 natural observer
time derivatives, one of which does not commute with index shifting
leading to 4 distinct time derivatives of the contravariant
and covariant forms of a vector field,
and therefore to 12 different acceleration 
decompositions, which double in number when one includes the 
Fermat's principle inspired optical conformal transformation 
variation of this decomposition. The utility of any of these 
24 possibilities for describing test particle motion in any 
spacetime remains to be justified in particular applications.

In the very special case of a static axisymmetric spacetime, the
various points of view and observer time derivatives compatible
with the symmetry each collapse from three to one in number, and
these 24 descriptions reduce to 2 in number, the standard
relative observer approach and its optical conformal variation.
Each of these seem useful in understanding certain aspects of the
qualitative behavior of test particle motion, and the latter is
clearly an elegant geometrization of the relative motion of
massive and massless test particles. 

The present article summarizes this situation in an attempt to
bring more understanding among the current groups  in this field 
of each other's distinct points of view.

\section{The Roster of Players}

A number of different approaches have been taken in studying the
properties of circular orbits in static and stationary 
spacetimes. One approach followed by De Felice, Semer\'ak, Page, 
and others attacks the problem from the spacetime point of view,
while others use a space-plus-time decomposition of the 
4-acceleration with respect to a family of test observers,
leading to an interpretation involving inertial forces due to
their motion.
Of the latter Abramowicz et al have been evolving a 
generalization of such an approach from the special static 
axisymmetric case up to general spacetimes, while Bini, Carini, 
and Jantzen have specialized gravitoelectromagnetism, the 
relativity of observer splitting formalisms, from the case of 
general spacetimes down to stationary and axisymmetric ones.
The tools of this last approach provide a necessary clarification 
and comparison of all the work in this area.
In particular the gradual reorientation of the Abramowicz et al
4-acceleration decomposition from the static case to general 
spacetimes is shown to have as its target the conformal 
modification inspired by Fermat's principle of the standard
space-plus-time splittings of the various observer congruence
approaches (congruence, threading, hypersurface, slicing) in use
today.

The players in the current
inertial forces / test particle motion game
fall into a number of groups.

\def\nvs{-6pt}

\vskip\nvs
\subsection*{Abramowicz et al
\cite{abrlas74}$^{\!-\,}$\cite{gupiyepra97}}
\vskip\nvs

Abramowicz and Lasota examined the roots of their later work in 1974 but
did not really begin studying inertial, especially centrifugal, forces
in static and stationary spacetimes until
Abramowicz and Lasota \cite{abrlas86} (1986)
and
Abramowicz, Carter, and Lasota \cite{abrcarlas} in 1988
(as an alternative to an earlier approach of De Felice, for example,
as they explicitly note),
and were then joined in continuing this work and extending it to
general spacetimes in a series of applications and
sequential corrections by
Prasanna,
Chakrabarti, 
Bi\v c\'ak,
Miller, Stuchl\'\i k,
Nurowski, Wex,
Sonego, Lanza, 
Massar,
and Iyer.

\vskip\nvs
\subsection*{Bini, Carini, and Jantzen
\cite{jancar91}$^{\!-\,}$\cite{embook}}
\vskip\nvs

Jantzen and Carini (1991) were later joined by Bini
in studying the larger picture of the 
relativity of spacetime splitting formalisms
and inertial forces in test particle motion,
which they called gravitoelectromagnetism.
More recently they have specialized this work to
some familiar stationary axisymmetric spacetimes.


\vskip\nvs
\subsection*{De Felice and Usseglio-Tomasset
\cite{def71}$^{\!-\,}$\cite{defuss96}}
\vskip\nvs

De Felice (1971) began by studying properties of 
test particle motion in black hole spacetimes
and did some preliminary work looking at the inertial force analogy
in 1975,
but did not use the inertial force language, instead
relying on the angular velocity variable to express
spacetime results for circular orbits.
With Usseglio-Tomasset in 1991
he began 
overlapping with the Abramowicz group
work, finally joining with Semer\'ak.


\vskip\nvs
\subsection*{Vishveshwara and Iyer
\cite{honschvis}$^{\!-\,}$\cite{acnsv}}
\vskip\nvs

Vishveshwara started with collaborators
Honig and Sch\"ucking (1974) and
Greene and Sch\"ucking (1975),
examining the stationary rest frame in black hole spacetimes
and was later joined by  
Iyer (1993)
in examining test particle motion in Killing submanifolds using the
spacetime Frenet-Serret formalism
and then with
Nayak made contact with the Abramowicz inertial force ideas.


\vskip\nvs
\subsection*{Semer\'ak
\cite{sem93}$^{\!-\,}$\cite{pag97}}
\vskip\nvs

Semer\'ak started by examining 
stationary observer families in black hole spacetimes in 1993, 
clearly making contact with the work of de Felice,
and then continued on to inertial forces
and test particle motion in general 
and in black hole spacetimes starting in 1995,
in the context of the Abramowicz et al
and Bini-Carini-Jantzen work.
(See also earlier work by
Tsoubelis, Economou, and Stoghianidis 
\cite{tsoecosto86,tsoecosto87}
on spinning test particles in
black hole spacetimes.)
Page continued studying properties of stationary
observer congruences in the same spirit.~\cite{pag97}

\vskip\nvs
\subsection*{Barrab\`es, Boisseau, and Israel
\cite{barboiisr}}
\vskip\nvs

Along the way Barrab\`es, Boisseau, and Israel (1995)
injected some useful perspectives on the issues
of inertial forces in black hole spacetimes from
the hypersurface point of view.

\vskip\nvs
\subsection*{Rindler and Perlick
\cite{rinper}}
\vskip\nvs

Rindler and Perlick (1990) introduced a nice analysis of circular
orbits in stationary axisymmetric spacetimes which proved useful
in some of the above work. This was followed by interesting
work by Perlick on Fermat's principle in general relativity
(see, e.g., M\o ller \cite{mol})
and related ideas. (See also Rindler's recent 
article.~\cite{rin97})

\bigskip

Most of this work using a relative observer analysis
has its roots in the literature dating back to
the fifties, although there seems to be somewhat of a literature
horizon effect in the bibliographies of 
the current generation of articles. An extensive bibliography
is given elsewhere.~\cite{jancarbin92}
Similar questions about properties of test particle motion
are also studied directly from the spacetime point of view
without direct reference to a test observer family as already
noted above.
This article will discuss the situation from the observer
point of view. 
The spacetime metric $g$ is assumed to have signature {\tt(-+++)}
and $||X|| = |g(X,X)|^{1/2}$ will denote the length of a vector
field, while $X^\flat$ and $X^\sharp$ will denote the totally
covariant and totally contravariant forms of a tensor field obtained
by index shifting with the spacetime metric.

\section{The Game: Observing Test Particle Motion and Spin Precession
Effects}

One starts with 
1) a congruence of test observer world lines with 
4-velocity field $u$ covering the region of interest in a given
spacetime and then analyzes the properties of 
2) individual test 
particle (timelike or null) world lines with 4-velocity $U$ (timelike
case) or 4-momentum $P$ (null case) defined only along each
such world line.

For a single such test world line, attention is thus confined to a
``{\bf relative observer world 2-sheet}"
in spacetime illustrated in Figure 1 
representing the
sheaf of observer world lines which cross the test particle world line.
This world sheet projects down to a curve in the quotient space of
observer world lines representing the trajectory in 
the test observer ``{\bf space}."
At each point of the test particle
world line, the tangent plane to this 2-sheet is
the ``{\bf relative observer 2-plane}" spanned by $u$ and $U$ or $u$ and
$P$ also containing the unit vectors representing the directions of 
relative motion.  
In the timelike case, for example, 
as illustrated in Figure 2, 
one has the reciprocal
pair of orthogonal decompositions 
\begin{eqnarray}
   U &=& \gamma(U,u) [ u + \nu(U,u) ] \ , \qquad 
           \hat\nu(U,u)=||\nu(U,u)||^{-1} \nu(U,u) 
          \nonumber\ ,\\
   u &=& \gamma(u,U) [ U + \nu(u,U) ] \ , \qquad 
           \hat\nu(u,U)=||\nu(u,U)||^{-1} \nu(u,U) \ ,
\end{eqnarray}
while in the null case one has only
\beq
  P = E(P,u) [ u + \hat\nu(P,u) ] \ ,
\eeq
where the hat notation indicates unit vectors.
One can also introduce the momentum representation of this
orthogonal decomposition for the timelike case (for convenience
set the rest mass $m=1$ so that $P=mU=U$)
\beq
  P = E(U,u) u + p(U,u) = \gamma(U,u) u + ||p(U,u)|| \hat\nu(U,u)  
      \ .
\eeq

\typeout{3 Pictex figures:}
\typeout{Loading Pictex macros:}

\input 
       prepictex
\input 
       pictex
\input 
       postpictex

\def\mathput#1{\relax \ifmmode \displaystyle #1\else 
$\displaystyle #1$\fi}


\typeout{Pictex Figure 1:}

\begin{figure}[hp]
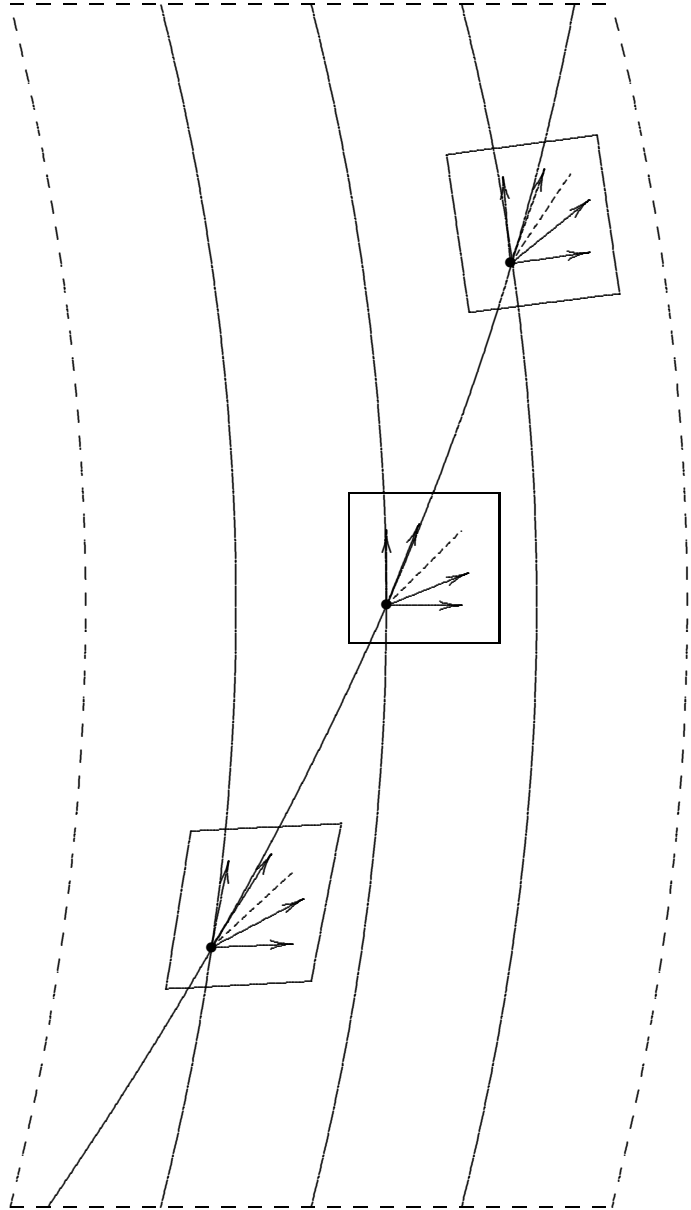
\label{fig:rows}
$$ \vbox{
\beginpicture
  \setcoordinatesystem units <1cm,1cm> point at 0 100  

    \putrule from  1.5 -0.5   to  1.5 1.5  
    \putrule from -0.5 -0.5   to  1.5 -0.5 
    \putrule from -0.5 -0.5   to -0.5 1.5  
    \putrule from -0.5  1.5   to  1.5 1.5  
  \setlinear
    \arrow <.3cm> [.1,.4] from  0 0  to  0 1 
    \arrow <.3cm> [.1,.4] from  0 0  to  1 0 
    
    \arrow <.3cm> [.1,.4] from 0 0  to 0.436  1.09    
    \arrow <.3cm> [.1,.4] from 0 0  to 1.09 0.436     

  \setdashes <2pt>
    \plot 0 0  1 1 /  
  \setsolid

    \plot  1.1 3.9    3.1 4.15 /      
    \plot  3.1 4.15     2.8 6.26 /  
    \plot  1.1 3.9    0.8 6 /  
    \plot  0.8 6     2.8 6.26 /  

    \arrow <.3cm> [.1,.4] from  1.65 4.55  to  1.55 5.7 
    \arrow <.3cm> [.1,.4] from  1.65 4.55  to  2.7 4.7 
  
    \arrow <.3cm> [.1,.4] from  1.65 4.55  to  2.1 5.8 
    \arrow <.3cm> [.1,.4] from 1.65 4.55  to 2.7 5.4     

  \setdashes <2pt>
    \plot 1.65 4.55 2.45 5.75 /  
  \setsolid

      \plot  -2.93 -5.1   -1 -5 / 
      \plot  -1 -5    -0.6 -2.9 / 
      \plot  -2.93 -5.1   -2.6 -3 /  
      \plot  -2.6 -3  -0.6 -2.9  /  
  \setlinear
    \arrow <.3cm> [.1,.4] from  -2.33 -4.55  to  -2.1 -3.4 
    \arrow <.3cm> [.1,.4] from  -2.33 -4.55  to  -1.25 -4.5 
    
  \arrow <.3cm> [.1,.4] from -2.33 -4.55  to -1.53 -3.3   
  \arrow <.3cm> [.1,.4] from -2.33 -4.55  to -1.1 -3.9  

  \setdashes <2pt>
    \plot -2.33 -4.55  -1.25 -3.55 /  
  \setsolid


\setquadratic

   \plot -1 -8  0 0  -1 8 /   
   \plot  1 -8  2 0   1 8 /   
   \plot -3 -8 -2 0  -3 8 /   
   \plot -4.5 -8 0 0  2.5 8 /   

\setdashes
   \plot  3 -8  4 0   3 8 /   
   \plot -5 -8 -4 0  -5 8 /   
   \putrule from -5 -8 to 3 -8 
   \putrule from -5  8 to 3  8 
\setsolid
\setlinear

 \put {\mathput{\bullet}}          at  0 0            
 \put {\mathput{\bullet}}          at  1.65 4.55      
 \put {\mathput{\bullet}}          at  -2.33 -4.55    


\endpicture}$$

\caption{
The relative observer world sheet of $u$ and $U$.}

\end{figure}

\typeout{Pictex Figure 2:}

\begin{figure}[hp]
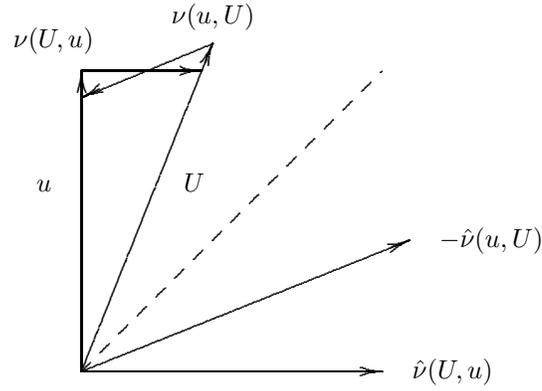
\label{fig:rop}
$$ \vbox{
\beginpicture
  \setcoordinatesystem units <0.8cm,0.8cm> point at 0 100  

    \putrule from 0 0   to 0 5.0 
    \putrule from 0 0   to 5.0 0  
    \putrule from 0 5.0 to 2.0 5.0 
    
  \setlinear
 
    \plot 0 0 2.18  5.45  /  
    \plot 0 0 5.45 2.18  /     
    \plot 0 4.55 2.18 5.45 / 
    
\setdashes
    \plot 0 0  5.0 5.0 /  

\setsolid

\setlinear

    \arrow <.3cm> [.1,.4]    from  0 4.7 to 0 5.0           
    \arrow <.3cm> [.1,.4]    from  2.06 5.15 to 2.18 5.45   

    \arrow <.3cm> [.1,.4]    from  1.7 5.0 to 2.0 5.0       
    \arrow <.3cm> [.1,.4]    from  0.3 4.7 to 0 4.55        
 
    \arrow <.3cm> [.1,.4]    from  4.7 0 to 5.0 0           
    \arrow <.3cm> [.1,.4]    from  5.15 2.06 to  5.45 2.18  
    

  \put {\mathput{u}}                          [rB]   at  -.5 3.0
  \put {\mathput{U}}                          [lB]   at  1.7 3.0

  \put {\mathput{\nu(U,u)}}                   [rB]   at  0.2 5.4
  \put {\mathput{\nu(u,U)}}                   [lB]   at  1.5 5.8

  \put {\mathput{\hat\nu(U,u)}}                   [l]   at  5.5 0.0
  \put {\mathput{-\hat\nu(u,U)}}                  [l]   at  5.95 2.18

\endpicture}$$

\caption{%
The relative
observer plane of $u$ and $U$. The unit vector
$\hat\nu(U,u)$ gives the direction of the longitudinal subspace of
the local rest space $LRS_u$ relative to $U$, while
$\hat\nu(u,U)$ does the same for $LRS_U$.
}

\end{figure}

\typeout{Pictex Figure 3:}

\begin{figure}[hb]
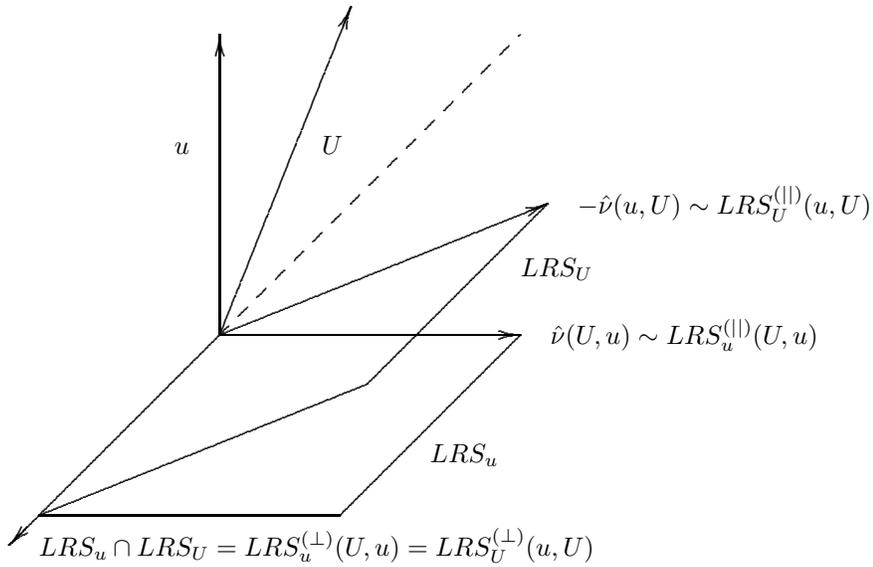
\label{fig:rmod}
$$ \vbox{
\beginpicture
  \setcoordinatesystem units <0.8cm,0.8cm> point at 0 100  

    \putrule from 0 0   to 0 5.0  
    \putrule from 0 0   to 5.0 0  
    \putrule from -3 -3   to 2.0 -3  

  \setlinear
 
    \plot 0 0 2.18  5.45  /       
    \plot 0 0 5.45 2.18  /        

    \plot 0 0 -3.5 -3.5 /         

    \plot -3 -3 2.45 -0.82  /     
    \plot 2.45 -0.82  5.45 2.18 /  

    \plot  2 -3  5 0 /            

\setdashes
    \plot 0 0  5.0 5.0 /  

\setsolid

\setlinear

    \arrow <.3cm> [.1,.4]    from  0 4.7 to 0 5.0           
    \arrow <.3cm> [.1,.4]    from  2.06 5.15 to 2.18 5.45   
 
    \arrow <.3cm> [.1,.4]    from  4.7 0 to 5.0 0           
    \arrow <.3cm> [.1,.4]    from  5.15 2.06 to  5.45 2.18  

    \arrow <.3cm> [.1,.4]    from  -3.2 -3.2 to -3.5 -3.5   


  \put {\mathput{u}}                          [rB]   at  -.5 3.0
  \put {\mathput{U}}                          [lB]   at  1.7 3.0

  \put {\mathput{LRS_u}}                      [l]   at  3.5 -2.0
  \put {\mathput{LRS_U}}                      [l]   at  5.0 1.09

  \put {\mathput{\hat\nu(U,u) \sim LRS_u^{(||)}(U,u)}}   [l] at  5.5 0.0
  \put {\mathput{-\hat\nu(u,U) \sim LRS_U^{(||)}(u,U)}}  [l] at  5.95 2.18

  \put {\mathput{LRS_u \cap LRS_U 
         = LRS_u^{(\bot)}(U,u) = LRS_U^{(\bot)}(u,U)}}   [l] at  -3.0 -3.5

\endpicture}$$

\caption{%
The relative motion orthogonal decomposition of the local rest spaces
of $u$ and $U$. The intersection of the two local  rest spaces
is the common 2-dimensional transverse subspace of
the tangent space.
}

\end{figure} 
\clearpage

In the timelike case one has a further 
``{\bf relative motion orthogonal decomposition}"
illustrated in Figure 3 
of the two local rest spaces $LRS_u$ and $LRS_U$ into the common 
2-dimensional subspace $LRS_u \cap LRS_U$ transverse to the 
direction of relative motion
and its 1-dimensional (longitudinal)
orthogonal complement along the direction of relative motion in each
such rest space. 
In the null case one has only the orthogonal
decomposition of $LRS_u$ into the 1-dimensional longitudinal
subspace along
the direction of relative motion and its 2-dimensional transverse
orthogonal complement.

Spin precession effects for a test particle are studied via a 
vector $S$ belonging to $LRS_U$ undergoing Fermi-Walker transport 
along the test world line in the timelike case. The behavior
of the world line itself as seen by the observer family requires
the choice of a 
geometric temporal derivative operator along the test world line, 
equivalent to the choice of a spatial frame along that world line
(having zero derivative),
in order to measure changes in the components of the measured fields
as they evolve along the world line, given a choice of parametrization
for it. A number of such geometric derivatives along a curve are
available, leading to collection of quantities that are defined by
derivative formulas, depending on the choice of this derivative.

This mathematical setting has a natural mathematical structure
which requires no invention, which is unambiguous, and 
which has a direct physical interpretation. The key concepts
are summarized in Table 1.

\begin{table}
\caption{Summary of the key concepts underlying the 
natural mathematical structure of this problem.}
\begin{itemize}
\item[]
\begin{tabular}{ll}
{\bf observer congruence:} &
orthogonal projection, \\
& kinematical quantities\\
\qquad $+$ & \\
{\bf test particle world line:} &
intrinsic derivative\\
\qquad 
$ \Longrightarrow$ & \\
{\bf relative observer world sheet:} &
projected intrinsic derivatives
\end{tabular}
\end{itemize}
\end{table}

\section{Inertial forces: 
Measuring the 4-Acceleration/4-Force Equation}

The 4-acceleration $a(U)=D U / d \tau_U$
of a timelike test particle parametrized by the
proper time $\tau_U$ is the intrinsic derivative
of the 4-velocity along its world line.
The equation of motion for the test particle is
the 4-acceleration/4-force equation
\beq
  a(U) = f(U) \ .
\eeq
This is measured by the observer family by 
orthogonally projecting it into
a family of spatial fields, in this case a scalar and
a spatial vector. The temporal projection along $u$
leads to the evolution equation for the observed
energy of the test particle along its world line, 
while the spatial projection 
orthogonal to $u$ leads to the evolution equation for
the observed 3-momentum of the particle along its
world line, with the kinematical quantities describing
the motion of the family of observers entering these
measured equations as ``{\bf inertial forces}."
The spatial projection 
operator $P(u)$ acts on a tensor by contraction of all 
indices with the corresponding projection
tensor, here identified with the same
symbol $P(u) = Id + u\otimes u^\flat$. The covariant form
of this mixed index tensor is the spatial metric
$h(u)=P(u)^\flat = P(u) g$.

Dividing the measured spatial projection equation
by the gamma factor of the test particle in
the timelike case, one may absorb this factor into
a reparametrization of the world line by the sequence
of observer proper times $\tau_{(U,u)}$
along it rather than by
the test particle's own proper time $\tau_U$, related by
$ d\tau_{(U,u)} / d\tau_U = \gamma(U,u) $.
The result
\beq
   \gamma(U,u)^{-1} P(u)  a(U) 
     = \gamma(U,u)^{-1} P(u) f(U)
\eeq
is directly equivalent to 
\beq
  D_{\rm(fw)}(U,u) \, p(U,u) / d \tau_{(U,u)} - 
F^{\rm(G)}_{\rm(fw)}(U,u) 
     = F(U,u) \ ,
\eeq
where $P(u) D / d\tau_U = D_{\rm(fw)}(U,u) / d\tau_U$ is the 
spatial Fermi-Walker intrinsic derivative along the test world line
and $F(U,u)=\gamma(U,u)^{-1}P(u)f(U)$ is the measured 3-force.
The term appearing with a minus sign on the left hand side of this
projected, rescaled 4-acceleration/4-force equation
\begin{eqnarray}
   F^{\rm(G)}_{\rm(fw)}(U,u)
              &=& - \gamma(U,u)^{-1} P(u) D \, u / d \tau_U 
               = - D_{\rm(fw)}(U,u) \, u / d \tau_{\rm(U,u)} 
                     \nonumber\\
              &=& \gamma(U,u) [ -a(u) 
                   + \{-\omega(u)+\theta(u)\} \rightcontract \nu(U,u) ]
                  \nonumber\\
              &=& \gamma(U,u) [ g(u) 
                          + H_{\rm(fw)}(u)\rightcontract \nu(U,u) ] \ ,
\end{eqnarray}
arising from the intrinsic derivative of $u$ along $U$ with respect to
the test particle proper time $\tau_U$, may be moved to the right hand
side with a positive sign in order to be interpreted as inertial forces 
due to the motion of the observers themselves.
These inertial forces involve the kinematical quantities of the
observer congruence, namely the acceleration vector field $a(u)$
(leading to a gravitoelectric field and a spatial 
gravitoelectric gravitational force) 
and their vorticity/rotation $\omega(u)$ and expansion $\theta(u)$
mixed tensor fields
(leading to a gravitomagnetic vector field and tensor field and a 
Coriolis or gravitomagnetic force linear in the relative velocity, 
modulo the gamma factor).
This reflects the observer measurement orthogonal decomposition at the
derivative level.

The further relative motion orthogonal decomposition of the observer
local rest spaces $LRS_u$ along the test particle world line may be
similarly used to decompose the temporal derivative of the 3-momentum
into a longitudinal relative acceleration term and a transverse one
\begin{eqnarray}
\fl\qquad
   p(U,u) &=& ||p(U,u)|| \hat\nu(U,u) \ , \nonumber\\
\fl\qquad
   D_{\rm(fw)}(U,u) \, p(U,u) / d \tau_{(U,u)} 
      &=& [ D_{\rm(fw)}(U,u) ||p(U,u)|| / d\tau_{(U,u)} ] \hat\nu(U,u) 
       \nonumber\\ 
\fl\qquad   
      &&  + ||p(U,u)|| \, 
           D_{\rm(fw)}(U,u) \, \hat\nu(U,u) / d\tau_{(U,u)}
            \ ,
\end{eqnarray}
the latter of which may be re-expressed in the form
\begin{eqnarray}
     [D_{\rm(fw)}(U,u) \, p(U,u) / d \tau_{(U,u)}]^{(\bot)}
        &=& \gamma(U,u) a_{\rm(fw)}^{(\bot)} (U,u)
               \ ,\nonumber\\
     a_{\rm(fw)}^{(\bot)} (U,u)
       &=& \nu(U,u)^2 
          [ D_{\rm(fw)}(U,u) \, \hat\nu(U,u) / d\ell_{(U,u)} ]
         \ , 
\end{eqnarray}
where the derivative has been reparametrized using the spatial 
arclength parametri\-zation, corresponding to the relationship
$d \ell_{(U,u)} / d\tau_{(U,u)} = ||\nu(U,u)||$,
and where 
$$ a_{\rm(fw)}(U,u)
= D_{\rm(fw)}(U,u) \, \nu(U,u) /d\tau_{(U,u)} $$
is the Fermi-Walker relative acceleration.
Apart from the additional gamma factor,
the second term in the decomposition of the spatial
momentum rate of change above is the transverse part
of the relative acceleration, namely the 
Fermi-Walker relative centripetal acceleration.
The factor in square brackets in the latter expression
may be expressed in terms of its 
direction (relative normal) and magnitude (relative curvature)
\begin{eqnarray}
  D_{\rm(fw)}(U,u) \, \hat\nu(U,u) / d \ell_{(U,u)} 
      &=& \kappa_{\rm(fw)}(U,u) \, \hat\eta_{\rm(fw)}(U,u)  \ ,
\end{eqnarray}
with the reciprocal of the magnitude defining the
Fermi-Walker relative curvature $\rho_{\rm(fw)}(U,u)$
of the world line, in terms of which the 
Fermi-Walker relative 
centripetal acceleration takes its usual 
form familiar from nonrelativistic mechanics
\beq
     a_{\rm(fw)}^{(\bot)} (U,u)
       = [ \nu(U,u)^2 / \rho_{\rm(fw)}(U,u) ] 
               \hat\eta_{\rm(fw)}(U,u)  \ .
\eeq
Continuing this leads to the 
Fermi-Walker relative bi-normal $\hat\xi_{\rm(fw)}(U,u)$
and torsion $ \tau_{\rm(fw)}(U,u)$
for the test world line and the remaining
relative Frenet-Serret equations
in complete analogy with the geometry
of a curve in a three-dimensional Riemannian 
manifold.~\cite{binjanmer97}
These remaining relative Frenet-Serret equations are given by
\begin{eqnarray}
   \frac{D_{\rm(fw)} (U,u)}{d\ell_{(U,u)}}\hat\eta_{\rm(fw)}(U,u) 
       & =& -k_{\rm(fw)}(U,u)\hat\nu(U,u)
            +\tau_{\rm(fw)} (U,u) \hat\xi_{\rm(fw)}(U,u)
          \ ,\nonumber \\
   \frac{D_{\rm(fw)} (U,u)}{d\ell_{(U,u)}}\hat\xi_{\rm(fw)}(U,u) 
       & =& -\tau_{\rm(fw)} (U,u) \hat\eta_{\rm(fw)}(U,u) \ ,
\label{relfs}
\end{eqnarray}
or
\beq
\frac{D_{\rm(fw)} (U,u)}{d\ell_{(U,u)}}E_{\rm(fw)}(U,u)_a
    =\omega_{\rm(fw)}(U,u)\times_u E_{\rm(fw)}(U,u)_a  \ ,
\eeq
where
\beq
\fl\qquad
  \{E_{\rm(fw)}(U,u)_a\}
     = \{ \hat\nu(U,u), \hat\eta_{\rm(fw)}(U,u), \xi_{\rm(fw)}(U,u)
=\hat\nu(U,u)\times_u \hat\eta_{\rm(fw)}(U,u) \}
\eeq
and
$\omega_{\rm(fw)}{(U,u)}
 = \tau_{\rm(fw)}(U,u)\hat\nu(U,u)
+k_{\rm(fw)}(U,u) \hat\xi_{\rm(fw)}(U,u)$.

The centripetal
acceleration term may also be moved to the right hand side of this
equation with a minus sign
to be interpreted as a ``{\bf centrifugal force}," which might be
useful for the case of relative motion at fixed speed where 
the longitudinal
acceleration vanishes (purely transverse relative motion), in order
to interpret the acceleration/force equation as a balance of spatial
forces. 
Although this transverse term also belongs to the test particle
local rest space $LRS_U$, it does not directly correspond to a force
in that frame, since it only has meaning in the context of the entire
observer local rest space, where the remaining terms in this 
projected acceleration/force equation live in general, and is also
scaled by a factor of gamma compared to a component of the 4-force
in the test particle local rest space.

The longitudinal 3-acceleration term may also be moved to the right
hand side with a minus sign and interpreted as an ``{\bf Euler}"
force in the terminology of Abramowicz et al, so that the entire
equation may be thought of as a balance of 3-forces from the point
of view of the family of observers along the test world line.

\section{Variations of the Projected Intrinsic Derivative}

Observer  measurements of tensor fields and tensor field differential
operators involve the observer measured derivative operators
\begin{eqnarray}
  \nabla(u) X &=& P(u) \nabla X\ , 
      \nonumber\\
  \nabla_{\rm(fw)}(u) X &=& P(u) \nabla_u X\ , \quad
  \nabla_{\rm(lie)}(u) X = P(u) \Lie_u X \equiv \Lie(u)_u X \ .
\end{eqnarray}
The first is the spatial covariant derivative, for which the spatial 
metric $h(u)$ is covariant constant.
In addition to the last two of these which are temporal derivatives,
namely the spatial Fermi-Walker derivative and the spatial 
Lie derivative along $u$, 
one may introduce a third temporal derivative, the corotating
spatial Fermi-Walker derivative, so that the three are related
as follows when acting on a vector field $X$
\begin{eqnarray}
      \nabla_{\rm(cfw)}(u) X
      &=& \nabla_{\rm(fw)}(u) X
           +   \omega(u) \rightcontract X
       \nonumber\\
      &=& \nabla_{\rm(lie)}(u) X
           +  \theta(u) \rightcontract X
     \ ,
\end{eqnarray}
differing among themselves only by the action of
a linear transformation of the observer local rest space. 
The spatial metric has vanishing derivative with respect to only
the Fermi-Walker and corotating Fermi-Walker such derivatives, while
it has a nonvanishing derivative 
$\nabla_{\rm(lie)}(u) \, h(u) = 2 \theta(u)^\flat$ so that
index-shifting with the spatial metric does not commute with this latter
derivative unless the expansion tensor vanishes, as occurs in 
the case of 
a stationary Killing observer congruence in a stationary spacetime,
where the Lie and corotating spatial temporal derivatives along
the observer congruence coincide.

If $X$ is a tensor field on spacetime, its intrinsic derivative
along a timelike test particle world line with 4-velocity $U$
equals its covariant derivative along $U$
\beq
   D X / d\tau_U = \nabla_U X 
     = \gamma(U,u) [ \nabla_u +\nabla_{\nu(U,u)} ] X \ .
\eeq
Spatially projecting this leads to the Fermi-Walker spatial 
intrinsic derivative along the world line. 
Changing the parametrization
to observer proper time as well leads to
\beq
   D_{\rm(fw)}(U,u) \, X / d\tau_{(U,u)}
     =  \nabla_{\rm(fw)}(u) X + \nabla(u)_{\nu(U,u)} X \ .
\eeq
This may be extended to define two other spatial intrinsic derivatives
along the test world line by replacing the Fermi-Walker temporal
derivative by the corotating one or the spatial Lie derivative along
$u$
\beq\fl\qquad
   D_{\rm(tem)} (U,u) \, X / d\tau_{(U,u)}
     =  \nabla_{\rm(tem)}(u) X + \nabla(u)_{\nu(U,u)} X \ ,
     \quad {\rm\scriptstyle tem  = fw,cfw,lie}\ ,
\eeq
which are then related to each other in the same way as the observer
temporal derivatives that they generalize. For example, for a
vector field $X$ one has
\begin{eqnarray}      
  D_{\rm(cfw)}(U,u) \, X / d \tau_{(U,u)}
      &=& D_{\rm(fw)}(U,u) \, X / d \tau_{(U,u)}
           +   \omega(u) \rightcontract X
       \nonumber\\
      &=& D_{\rm(lie)}(U,u) \, X / d \tau_{(U,u)} 
           +  \theta(u) \rightcontract X
     \ ,
\end{eqnarray}

In order to express these operators in a form in which they may
be directly applied to a field defined only along the world line,
one obtains a formula analogous to the one for the intrinsic
derivative itself with respect to some choice of spacetime
frame, involving
the components of the connection with respect to that frame
\beq
   D X^\alpha / d \tau_U
      = d X^\alpha / d \tau_U 
           + \Gamma^\alpha{}_{\beta\gamma} U^\beta X^\gamma \ .
\eeq
For the projected intrinsic derivatives, reparametrized to
observer proper time, the appropriate projections of these
connection component terms are joined by additional
kinematical terms in the ``cfw" and ``lie" cases
\beq\fl 
   D_{\rm(tem)} \, X^\alpha / d \tau_{(U,u)}
      = d X^\alpha / d \tau_{(U,u)} 
           + \Delta_{\rm(tem)}(u)^\alpha{}_{\beta\gamma} 
                     U^\beta X^\gamma / \gamma(U,u) \ , 
     \  {\rm\scriptstyle tem  = fw,cfw,lie}\ . 
\eeq
The projections of the connection component terms themselves involve
the spatial connection components (associated with the spatial metric
$h(u)$) and kinematical terms. However,
only the Fermi-Walker and corotating Fermi-Walker such derivatives
commute with index-shifting, while the Lie such derivative of the spatial
metric is
\beq
    D_{\rm(lie)}(U,u) \, h(u) / d\tau_{(U,u)}
        = 2 \theta(u)^\flat \ ,
\eeq
so an additional expansion term appears when shifting indices on an
object being differentiated by this derivative, so that derivatives
of the contravariant and covariant forms of a vector field differ
by an expansion term. 
Another consequence of this
is that its derivative of a unit spatial vector field is not orthogonal
to the original vector field in general
\beq\fl\quad
   \hat\nu(U,u) \cdot D_{\rm(lie)}(U,u) \, \hat\nu(U,u) / d\tau_{(U,u)} 
         = - ||\nu(U,u)||^{-1}
                   \theta(u)^\flat (\hat\nu(U,u),\hat\nu(U,u)) \ ,
\eeq
in contrast with the situation for the other two derivatives which
respect spatial inner products.

Thus for a given observer congruence, one has four different natural
``{\bf relative Frenet-Serret structures}"
along a test particle world line
corresponding to the three temporal derivative operators with
the index-shifting complication, and four
different decompositions of the acceleration/force equation
and its respective terms, corresponding to different pairs of
numerical coefficients of the vorticity and expansion terms in
the gravitomagnetic field tensor.

For a stationary spacetime with a stationary observer congruence,
it is most natural to use the Lie spatial intrinsic derivative,
since it is the one most closely connected to the spatial geometry
without additional kinematical linear transformations of the
spatial tangent space. If the observer congruence is along
a Killing direction, then this derivative
also commutes with index-shifting by the spatial metric,
and the Lie relative Frenet-Serret structure on the observer
quotient manifold coincides with the Riemannian
Frenet-Serret structure of the time-independent spatial metric
there.
If the relative observer world sheet is a group orbit in
the stationary axisymmetric case (circular orbits), 
then the Lie spatial intrinsic
derivative of the unit spatial velocity is also transverse.

The most interesting nontrivial case to study along these lines
is that of the (noncircular) spherical accelerated orbits 
and geodesics in a Kerr black hole spacetime.
These do not follow Killing trajectories and in general 
are not characterized by transverse relative acceleration.

\section{Further Complications}

Usually spacetime splittings are facilitated by the use of
adapted coordinate systems, or more generally by equivalence
classes of such coordinate systems which define a 
parametrized nonlinear reference frame. This ``{\bf full splitting
of spacetime}," as opposed to the partial splitting provided by
an observer congruence alone,
involves the 
independent structures of a slicing of the spacetime by a
family of hypersurfaces parametrized by a time function
and a threading of this family by a transversal congruence
of curves. Depending on the causality properties of these
two components one can adopt one of two possible points of view
using this structure.

If the congruence is timelike, one can interpret it as the
observer congruence, leading to the threading point of view.
If the slicing is spacelike, one can interpret its
field of unit normals as the
observer congruence, leading to the hypersurface point of view.
One can also adopt the slicing point of view using the threading
to evolve the fields in spacetime rather than the observer congruence
itself. All the geometry of the threading and hypersurface points
of view is identical with that already discussed above for the
observer congruence, while some new features arise from the
bi-conguence approach of the slicing point of view. In particular
the slicing point of view leads to a further temporal derivative
along the threading and along the test world lines.
This slicing point of view is the foundation of 
{\it Black Holes: The Membrane Paradigm\/} by Thorne et 
al.~\cite{thoetal} 
For black holes, the natural parametrized
nonlinear reference frame is the one associated with Boyer-Lindquist
coordinates, for which the zero-angular-momentum observers (ZAMO's)
are the hypersurface normal observers, the static Killing observers
are the threading observers, and the threading evolution vector field
is the time coordinate derivative Killing vector field. For any 
parametrized nonlinear reference frame, one has $4 \times 3 = 12$
(hypersurface, threading, slicing contravariant, slicing 
covariant and ``lie," ``cfw," ``fw") different 
decompositions of the 4-acceleration/4-force equation.

\section{Spatial Conformal Transformations}

For stationary spacetimes with a stationary family of observers
and a stationary parametrized
nonlinear reference frame (Lie dragged into itself
by the symmetry group action)
one may perform a conformal transformation
of the spatial geometry on the observer quotient space which absorbs
the spatial gravitational force term for massless test particles into
the projected intrinsic derivative, leaving only the Coriolis or
gravitomagnetic force remaining. (This can be done in any of the three
points of view: threading, hypersurface, and slicing,
although without stationarity, the part of the gravitoelectric
force involving the Lie derivative of the shift remains present as 
well. This doubles the number of variations of the force decomposition
to 24 for a parametrized nonlinear reference frame.)
This gravitomagnetic force vanishes in a static spacetime 
with a static Killing observer congruence (for which all 12
variations of the acceleration decomposition coincide)
so that the projection of null geodesics onto the observer quotient
space leads to geodesics of the conformally rescaled spatial geometry,
a fact known as Fermat's principle, well known for many decades.

Abramowicz et al have called this the 
``{\bf optical geometry}"  and Perlick the ``{\bf Fermat geometry}."
This change in variables essentially
shuffles terms in the measured acceleration/force equation between
the 3-acceleration (longitudinal and transverse terms)
and the spatial gravitoelectric force,
apart from conformal rescalings.

In a stationary spacetime, the optical geometry is not itself directly
linked to the paths of null geodesics, but in some sense to the average
motion obtained by averaging the gravitomagnetic deflection in opposite
directions of photon trajectories from the spatial geodesic paths.
It is yet to be seen how this can
help understand the physical properties of such spacetimes
or of more general ones.

\section{Difficulties from Not Using the Natural
Mathematical Structure}

\subsection{Angular Velocity versus 3-Velocity in Stationary
     Axisymmetric Spacetimes}

A minor inconvenience of the use of the angular velocity
(connected with global behavior)
to parametrize the description of local physical quantities
is that one has no sense of the magnitude of the physical
velocities involved, and graphs of various affects are distorted
by the map between the (coordinate) angular velocity
and angular physical component of the spatial velocity.
By using instead the latter physical component with respect
to natural observer families,
comparisons are more clear, especially when one is interested
in those velocities which correspond to timelike motion: the
velocities which bound this interval have absolute value 1,
while the corresponding bounding
angular velocities are in general
two complicated functions of the radius. However,
since for example, De Felice or Semer\'ak or Page
do not directly refer to observer congruences
but work with spacetime quantities, the
angular velocity variable is natural for them.
On the other hand, even in the spacetime discussion,
the ZAMO's provide a
natural way to convert angular velocities into
physical spatial velocities in the angular direction in
the entire exterior region of a black hole, and this
new choice of variable can
help better visualize the meaning of many graphs
frequently displayed in their analyses.

In particular, the radial acceleration of the often studied 
equatorial plane circular orbits in a rotating black hole 
spacetime
is a fractional quadratic function of the angular velocity
$\Omega$,
or of de Felice's fractional linearly related $y$ variable,
or of the physical velocity in the angular direction 
$\nu$ with
respect to any observers in circular motion, all variables
which
are related to each other by at most linear or fractional
linear transformations. The simplest and most directly
interpretable of all these variables seems to be the
physical velocity in the angular direction with respect to
the ZAMO observers, which are defined in the entire exterior
of the hole.~\cite{barboiisr,bincarjan97b} 
Its use 
rectangularizes the region in the radius-velocity plane
corresponding to timelike or lightlike motion
(for which $\nu\in [-1,1]$), removing the radial 
deformation of this region in the plane of the radius and
an angular velocity related variable.

\subsection{Abramowicz et al and Spatial Projections}

The fundamental equations of determining the observer
congruence in the Abramo\-wicz approach are now \cite{abrlas97}
\beq
   g(n,n) = -1\ , \quad
   \omega(n)=0\ ,  \quad
   a(n)= \nabla(n) \Phi \ .
\eeq
Given any unit hypersurface-forming timelike vector field $n$
on spacetime (the content of the first two equations),
the third equation defines an acceleration
potential for this observer congruence
(discussed e.g., by Ehlers \cite{ehl})
which is equivalent to a choice of lapse function
$N$ for the corresponding family of integral hypersurfaces of
$n$, or equivalently
a choice of time function $t$ parametrizing this
family, from which one may define the acceleration
potential $\Phi = -\ln dt(n)$ (equivalent to 
$N = e^\Phi$).

Omitting the spatial projection of the gradient in the definition
of the acceleration potential leads to the original version
of the third Abramowicz fundamental equation
\beq
     a(n) = \nabla \Phi\ .
\eeq
Since $0 = n \cdot a(n) = \nabla_n \Phi$,
it is clear that one must now impose
the consistency condition $ \Lie_n \Phi = 0 $
(equivalently $\Lie_n N$ = 0)
on the acceleration potential and observer 4-velocity
which limits the generality of the observer congruence.
This older version only permits observer
congruences for which the normal component 
$ n \cdot X$ of any
generating vector field
$ X $ for a time function $t$ parametrizing its
family of integral hypersurfaces ($ dt(X)=1$)
is Lie dragged along the congruence.
In  particular inertial observers of the Newtonian limit 
do not belong to the solution space of this older
version of the fundamental equations as noted by
Sonego and Massar,~\cite{sonmas} 
while they are clearly permitted
by the new fundamental equations which have an
arbitrary timelike congruence as their solution.

There is also the question of the missing spatial 
projections in the decomposition of the spatial
acceleration also noted by Sonego and Massar.
In various discussions of the approach, the 
``{\bf comoving
frame of the test particle}" (namely $LRS_U$)
has been brought into the
discussion of the various terms in the decomposition of
its 4-acceleration. However, in general only the
centripetal acceleration term, however defined, lies
in this local rest space and the observer local rest space.
The longitudinal acceleration,
however defined, lies in the observer local rest space,
while the gravitational force, however defined, need not
belong to either one unless spatially projected.
The only consistent interpretation of the various
force terms is to
spatially project them into the local rest space of the
observer. With the remarks of Sonego and Massar about
the Newtonian limit, this spatial projection has finally
been incorporated by Abramowicz et al.~\cite{abrlas97}

\subsection{Unnatural derivatives}

The use of an unnatural representation of derivatives along
the test particle world line instead of one of the projected
spatial intrinsic derivatives
obscures all of the calculations
of the Abramowicz et al approach and others like Semer\'ak
who have adopted the same notation when decomposing the
4-acceleration of a test particle world line.
These derivatives had their
origins in studying world lines which were integral curves
of Killing vector fields in stationary spacetimes, and so one
could trivially extend vectors along the world lines to the
entire relative observer world sheet and interpret derivatives
along the direction of relative motion of these extended 
fields. This does not work for more general world lines in
stationary spacetimes, and it requires detailed analysis
to understand what preferred extensions simplify the 
resulting derivative.

Although
formally ambiguous, one can almost guess how they were interpreted
in their actual calculations. Later they realized this
ambiguity when trying to extend their work to nonstationary
spacetimes
and tried to correct it by choosing a gauge condition
to extend the relevant fields along the world line to the entire
world sheet, but even this has had its problems
complicated by the fact that sign problems
plagued the inequivalent contravariant and covariant versions
of this gauge condition which alternately appear in different
discussions (index shifting does not commute with the 
Lie derivative).

Their intuition was clouded by the special features of 
transverse relative motion (circular orbits in stationary
axisymmetric spacetimes), so the need for an additional
spatial projection was not immediately understood. The same
remains true of their current gauge condition, 
which is still not the natural one for a general spacetime.

They use a gauge condition to extend the
spatial velocity off the test world line in order to define
their optical spatial derivative
\beq
  \tilde\nabla(n)_{\tilde{\hat\nu}(U,n)} \, \tilde{\hat\nu}(U,n)
\eeq
of the optical unit spatial velocity on the
test world line in a direction not tangent to the world line
on which it is defined.
Interpreting how this gauge condition relates their formal
derivatives to the various projected intrinsic derivatives
is a formidable barrier to follow their calculations,
especially since it seems to be
a moving target in the Abramowicz et al force decomposition
discussion.

The observer orthogonal decomposition of the
Lie derivative along the rescaled observer 4-velocity field
of a vector field $X$
(orthogonally decomposed as $X=X^\top(n) + X(n)$) is
\beq
    N^{-1} \Lie_{Nn} X 
       = [\Lie_n X^\top(n) ] n
                + \Lie(n)_n X(n) \ .
\eeq
In order to give meaning to the spatial derivative $\nabla(n)$
of the spatial velocity unit vector $\hat\nu(U,n)$ of the test 
particle, the gauge condition of Lie dragging along $Nn$, 
namely $\Lie_{Nn} \hat\nu(U,n) = 0$, which implies 
\beq
  \Lie(n)_n \hat\nu(U,n)
   =\nabla_{\rm(fw)}(n) \hat\nu(U,n)
        -\theta(n)\rightcontract \hat\nu(U,n) 
   = 0
\eeq
is used to extend to spatial velocity
from the world line to the entire observer world sheet containing
it. In fact this should only be done locally in some neighborhood
of the world line since a world line
which does not have unique intersections with each observer world
line it meets cannot in general be compatible with such a global
extension.

However, the more natural gauge condition on the local extension of
the (contravariant) optical unit spatial velocity
\begin{eqnarray}
&&   \nabla_{\rm(lie)}(n) \tilde{\hat\nu}(U,n) 
    = \Lie(n)_n \tilde{\hat\nu}(U,n) =0 
\Rightarrow \nonumber\\
&&   \tilde D_{\rm(lie)}(U,n) \, \tilde{\hat\nu}(U,n)^\beta
                                     / d \tilde\ell_{(U,n)}
         = \tilde\nabla(u)_{\tilde{\hat\nu}(U,n)} \,
                       \tilde{\hat\nu}(U,n)
\end{eqnarray}
makes their spatial derivative of this quantity agree with the
contravariant version of the 
natural optical spatial Lie intrinsic derivative 
with respect to the observer optical 
spatial arc length parametrization
(see (A.20) of Bini et al~\cite{bincarjan97a})
\begin{eqnarray}
  && \tilde D_{\rm(lie)}(U,n) \, \tilde{\hat\nu}(U,n)^\beta
                                     / d \tilde\ell_{(U,n)}
          \nonumber\\
  &&\quad = [ 1 / \tilde\nu(U,n) ]
       [ \nabla_{\rm(lie)}(n)
         + \tilde\nu(U,n) \tilde{\hat\nu}(U,n)^\alpha
                   \tilde\nabla(n)_\alpha ]
                       \tilde{\hat\nu}(U,n)^\beta
  \nonumber\\
  &&\quad  = N^2 [
     D_{\rm(lie)}(U,n) \, \hat\nu(U,n)^\beta / d \ell_{(U,n)}
  \nonumber\\
  &&\quad\ + P_n(U,n)^{(\bot)}{}^{\beta}{}_{\alpha}
                              \nabla(n)^\alpha \ln N
              + \hat\nu(U,n)^\beta / \nu(U,n) \,
                              \nabla_{\rm (lie)}(n) \ln N ]
   \ ,
\end{eqnarray}
where the first equality
defines the equivalent action of the new derivative on a congruence of
test particle world lines, while the second defines it for a single
such world line, and $\tilde\nabla(n)$ is the optical spatial
covariant derivative 
satisfying $\tilde\nabla(n) \tilde h(n) = 0$.
Their current
condition leads to additional terms which only vanish for
motion along Killing directions in a stationary spacetime
referred to a stationary observer congruence.
With our present choice of gauge condition, their final
decomposition of forces becomes the hypersurface point of
view optical decomposition (15.7) of Bini et 
al,~\cite{bincarjan97a} with their
Coriolis force term coinciding with the gravitomagnetic
term which in this case is due to the expansion
of the observer congruence alone.

Following their steps (and continuing revisions)
requires some patience, but the end
result is clear: their goal is just one of the 12 optical
possibilities
outlined above in the more general setting 
(including observer 4-velocity fields with nonzero vorticity
and the bi-congruence slicing point of view)
that follow
cleanly from performing a space-plus-time split of spacetime,
accompanied by the natural Fermat's principle
spatial conformal transformation.

\section{Cumulative Drag Index}

Prasanna and Iyer \cite{pra97,praiye97} have introduced a
``{\bf cumulative drag index}" for circular orbits
in the equatorial plane of the Kerr spacetime in an effort to
find some intrinsic characterization of spacetimes with
rotation.
In their notation for the optical decomposition of the radial
acceleration of such a circular orbit, this index is
\beq
  {\cal C} = \frac{Cf+Co-Gr}{Cf+Co+Gr}
\eeq
or
\beq
  {(1-{\cal C})/2} = \frac{Gr}{Cf+Co+Gr} \ ,
\eeq
where $Co$, $Cf$, and $Gr$ are the centrifugal
force, optical Coriolis force, and optical gravitational
force in the hypersurface point of view, modulo sign
(the acceleration term versus inertial force sign).
This may be rewritten in the gravitoelectromagnetism notation
in terms of orthonormal components with respect to the
orthogonal Boyer-Lindquist spatial coordinate frame 
(using the associated parametrized nonlinear reference frame) as
\begin{eqnarray}
  {(1-{\cal C})/2} &=& \frac{g(n)^{\hat r}}{a(U)^{\hat r}}
                 = \frac{{\cal F}(0;\kappa,\nu_+,\nu_-)}
                     {{\cal F}(\nu;\kappa,\nu_+,\nu_-)}
            \nonumber \\
     &=& \frac{ \nu_+ \nu_- (1-\nu^2)}{(\nu-\nu_-)(\nu-\nu_+)}
          \ ,
\end{eqnarray}
where $\kappa = \kappa(U,n)^{\hat\phi}$,
$\nu_\pm = \nu(U_\pm,n)^{\hat\phi}$,
$\nu = \nu(U,n)^{\hat\phi}$,
and $U_\pm$ are the geodesic 4-velocities
and where the physical component of the radial acceleration is
\begin{eqnarray}
  a(U)^{\hat r} &=& {{\cal F}(\nu;\kappa,\nu_+,\nu_-)}
                 = \kappa \frac{(\nu-\nu_-)(\nu-\nu_+)}{(1-\nu^2)}
            \nonumber \\
     &=& \gamma^2 [ \kappa \nu^2 - g(n)^{\hat r} 
                 - H_{\rm(fw)}(n)^{\hat r}{}_{\hat\phi} \nu ] \ .
\end{eqnarray}

The zero's of the denominator of the quantity $(1-{\cal C})/2$
coincide with those of ${\cal C}$ and occur at the geodesic
velocities. The zero's of the numerator of this quantity are
qualitatively similar to the zeros of ${\cal C}$ 
but in contrast have geometric meaning as the light velocities.
The gamma factor $\gamma^{-2}= 1-\nu^2$ leading to those zeros
may be removed from this quantity by instead taking the ratio of the
part of the total acceleration due to the gravitoelectric force and
the total acceleration itself
\beq
     \gamma^2 {(1-{\cal C})/2} 
         = \frac{-\gamma^2 g(n)^{\hat r}}{a(U)^{\hat r}}
         = \frac{\nu_-\nu_+}{(\nu-\nu_-)(\nu-\nu_+)} \ .
\eeq
One factor of gamma in this formula is part of the spatial 
gravitational force, while the second factor corresponds to the
conversion back to the test particle proper time force component,
in order to be compared to the test particle 3-acceleration.
In a similar way one could also introduce the ratios of the
gravitomagnetic or centripetal force terms to the total radial
acceleration, yielding a total of three new indices measuring
the relative importance of these three contributions to the
total acceleration.
However, but they are all infinite for geodesics
and do not seem to have a particularly intrinsic meaning
apart from the fact that the ZAMO observers are geometrically
defined by the symmetry of the spacetime.

\section{Towards a Better Understanding of the Arena of
Classical General Relativity}

We have no interest in attacking personalities or their work.
The fact that after all these years, such confusion can still
prevail in a topic like this shows the importance of clarifying
the basic tools we are using. All the formalism of 
gravitoelectromagnetism, the ``relativity of spacetime splittings,"
is not our invention, nor that of all the others who have worked
on various aspects of it over the past half century.
(References are given elsewhere.~\cite{jancarbin92})
It flows
naturally from the mathematical foundations of general relativity.
One only has to respect the natural mathematical structure already
present, and use a notation which unambiguously
describes all the possible variations that are allowed by that
structure. The utility of any of the approaches which fall under
this umbrella depends on the particular
application and what advantages in
understanding it can convey to us. Certainly introducing complication
in an existing spacetime which does not help us appreciate its
properties better than a direct spacetime approach is not useful,
but in many situations, a space-plus-time perspective does yield
useful information.

We appreciate all of the work done on this topic
in static spacetimes, where
the elegance of the application clearly provides the motivation
to extend it to stationary and more general spacetimes.
Indeed one may extend the optical geometry approach in a consistent
way not only to arbitrary hypersurface-forming timelike vector
fields in any spacetime, but to any arbitrary timelike vector field,
or to the context of the slicing approach of ADM. However,
unless we are able to communicate with each other in a clear way,
any advantages of one approach or another will remain locked in
that approach without the possibility of others appreciating its
meaning or translating it into their language.
Our ultimate goal after all is to shed more light on a rather
complicated theory, in a way that we can all benefit.
We hope that 
the ideas described in the present article
may contribute to a
better understanding among us about what we are all doing
and what it all means in relationship to each of our perspectives.

\typeout{references:}

\end{document}